\documentclass[iicol,pdflatex,sn-mathphys-num]{sn-jnl}
\usepackage{graphicx}%
\usepackage{multirow}%
\usepackage{amsmath,amssymb,amsfonts}%
\usepackage{amsthm}%
\usepackage{mathrsfs}%
\usepackage[title]{appendix}%
\usepackage{xcolor}%
\usepackage{textcomp}%
\usepackage{manyfoot}%
\usepackage{booktabs}%
\usepackage{algorithm}%
\usepackage{algorithmicx}%
\usepackage{algpseudocode}%
\usepackage{listings}%
\usepackage{nicematrix}
\usepackage{rotating}
\newcommand\tabrotate[1]{\begin{turn}{90}\rlap{#1}\end{turn}}

\begin{document}

\title[What exactly did the Transformer learn]{What exactly did the Transformer learn from our physics data?}

\author*[1]{\fnm{Martin} \sur{Erdmann}}\email{erdmann@physik.rwth-aachen.de}

\author[1]{\fnm{Niklas} \sur{Langner}}\email{niklas.langner@rwth-aachen.de}

\author[1]{\fnm{Josina} \sur{Schulte}}\email{josina.schulte@rwth-aachen.de}

\author[1]{\fnm{Dominik} \sur{Wirtz}}\email{dominik.wirtz2@rwth-aachen.de}

\affil[1]{\orgdiv{Physics Institute 3A}, \orgname{RWTH Aachen University},
\orgaddress{\street{Otto-Blumenthal-Stra{\ss}e}, \city{Aachen}, \postcode{52074}, %\state{state}, 
\country{Germany}}}

\abstract{Transformer networks excel in scientific applications. We explore two scenarios in ultra-high-energy cosmic ray simulations to examine what these network architectures learn. First, we investigate the trained positional encodings in air showers which are azimuthally symmetric. Second, we visualize the attention values assigned to cosmic particles originating from a galaxy catalog. In both cases, the Transformers learn plausible, physically meaningful features.
}




\maketitle

\section{Introduction}
\label{sec:introduction}

With the initial implementation of the Transformer, it was found to deliver excellent benchmark results for text translation tasks \cite{Vaswani:2017lxt}. This was followed shortly afterwards by successful applications of the Transformer to image recognition \cite{Dosovitskiy:2020qjv} and then as generative models with great impact when opened to the public \cite{openai_chatgpt_2022}. Due to its superior performance, the Transformer became established in science too and has replaced earlier approaches in a number of physics applications \cite{Qu:2022mxj,Finke:2023veq,PierreAuger:2023tyq,Wu:2024thh,Brehmer:2024yqw}. For a collection on Transformer models developed in particle physics refer to \cite{Feickert:2021ajf}.

For text- and image-based challenges, understanding the very good performance has been analysed in two aspects: 1) Positional encoding, that is information about arrangement patterns in the input data. 2) Attention, that concerns evaluation of correlations between input information. Both are convincingly demonstrated in self-attentions and translations, where annotated texts show that, after appropriate training, the Transformer evaluates connections between word sequences largely correctly. 

In scientific applications, the performance of Transformers is systematically evaluated by means of accuracy analyses or receiver operating characteristic (ROC) curves. However, it is often challenging to understand why Transformers can achieve outstanding performance. Recently, for a jet application in particle collisions, attention heat maps and particle-pair correlations have been used to demonstrate that the network learns traditional jet substructure observables \cite{Wang:2024rup}.

In this paper we show two applications from the physics of the highest-energy cosmic rays, where we want to understand through targeted visualizations what the excellent performance of the Transformers is based on. 

\begin{figure}[h]
\centering
\includegraphics[width=0.5\textwidth]{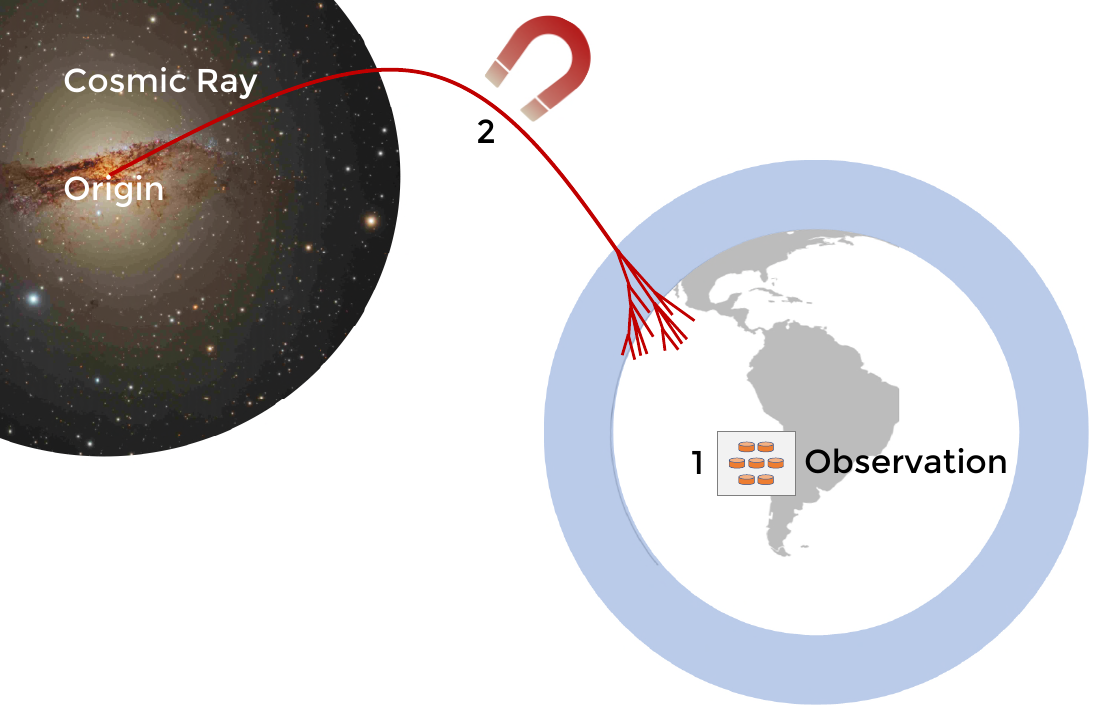}
\caption{ultra-high-energy cosmic rays propagate from their origin to an observatory on Earth. Two different Transformer networks are investigated here, first the hexagonal arrangement of the observatory's sensors, and second the displacement of the cosmic rays owing to magnetic fields.}
\label{fig:CR_Challenges}
\end{figure}

The scientific applications have two different objectives visualized in Fig.~\ref{fig:CR_Challenges}. The first is to investigate about exploiting the azimuthal symmetry of a hexagon-shaped sensor array. Here we focus on the aspect of positional encoding. The second case is about the assignment of cosmic particles to given galaxies or to background. Here we investigate the attentions of the particles. Both studies are based on simulation data.

\section{Positional encoding}

\paragraph{Scientific context}
ultra-high-energy cosmic rays are very likely composed of ionized atomic nuclei with masses ranging from protons up to iron. These cosmic particles are not observed directly but rather indirectly, through collisions with atoms in the Earth’s atmosphere (Fig.~\ref{fig:CR_Challenges}). These collisions generate so-called air showers, comprising billions of secondary particles. Detection of these air showers relies on a calorimetric measurement in which the Earth’s atmosphere serves as the absorber material, and readout is performed by a single layer of sensors at ground level. This setup is inherently stochastic: the relative detector coverage on the ground is minimal (on the order of \(10^{-5}\)), yet this low coverage remains fully sufficient because of the large number of secondary particles produced.

In the single readout layer, sensors are typically arranged in a hexagonal pattern (Fig.~\ref{fig:CR_Challenges}). Each sensor records both the temporal profile of incoming secondary particles and the signal amplitudes. To characterize the kinematics of the cosmic particles, information from all sensors is merged. Quantities of the cosmic particles of interest are: energy, arrival direction (represented by zenith and azimuth), and mass-related information through the shower depth and the muon count. While direction and energy can successfully be reconstructed with rule-based algorithms, the shower profile and muon count—which are key for determining the mass—are quantities where connectionist methods such as deep neural networks (e.g., Transformers) show potential advantages over rule-based approaches.

Details of how these observables are determined can be found in publications of the Pierre Auger Observatory \cite{PierreAuger:2015eyc,PierreAuger:2021fkf,PierreAuger:2021nsq,PierreAuger:2023tyq,PierreAuger:2024nzw}. 

\paragraph{Research on Transformer}
In this work, we focus on whether a Transformer, through its training process, exploits the symmetry of the hexagonal sensor arrangement to reconstruct cosmic-ray characteristics. Air shower physics is essentially rotationally symmetric with respect to the azimuth of the cosmic-ray arrival direction, although small deviations may arise—for instance, due to the Earth’s magnetic field. Rotational symmetry in azimuth can be captured by neural network architectures that employ so-called \textit{hexaconv} convolutions \cite{Hoogeboom:2018exn}, in which filters are rotated six times in accordance with the hexagonal geometry. By contrast, the Transformer architecture presented here does not include any explicit information about symmetries. However, the Transformer could in principle encode domain-specific symmetries contained in the training data within its positional encodings.

\begin{figure}[h]
\centering
\includegraphics[width=0.47\textwidth]{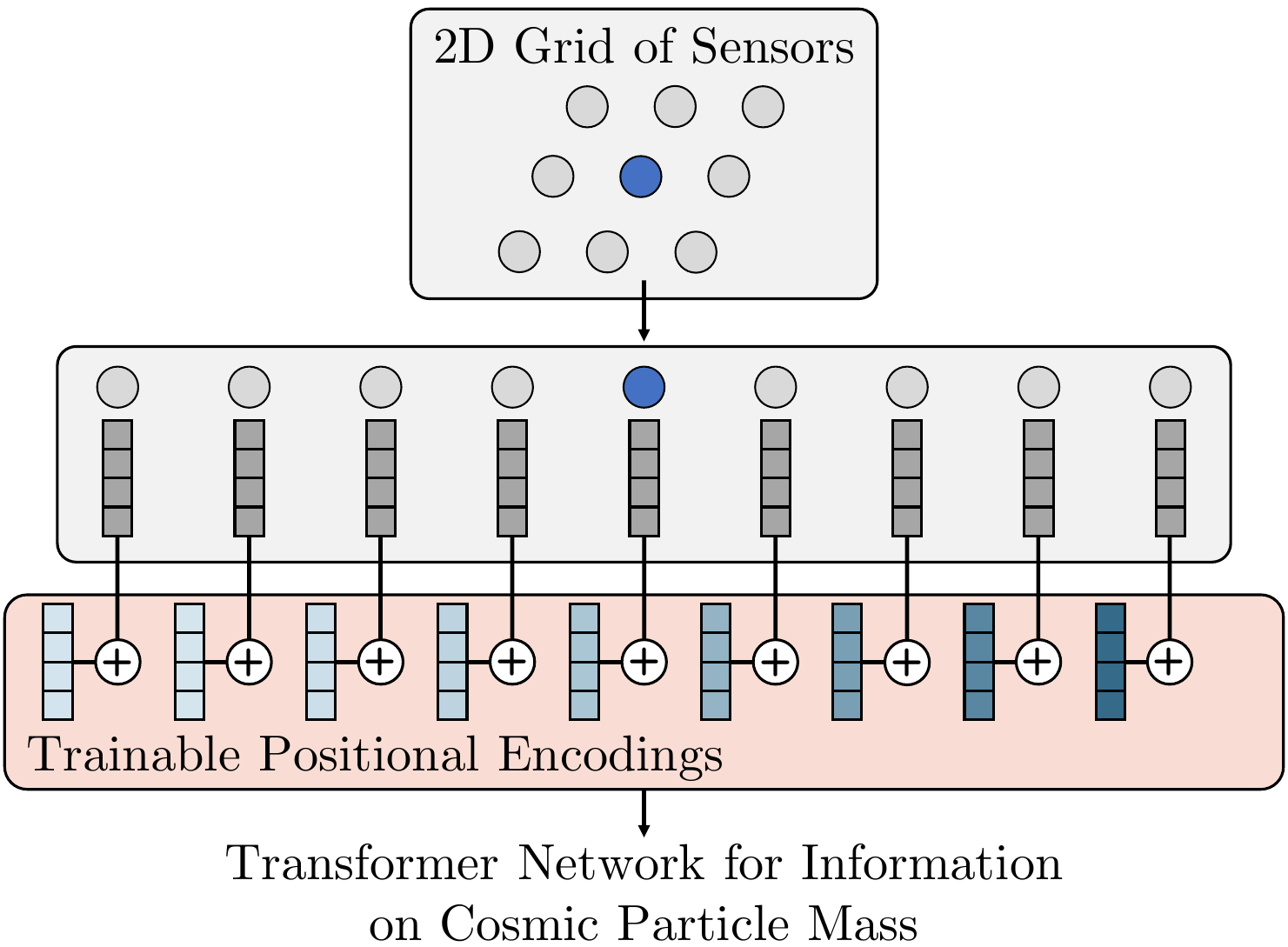}
\caption{Positional Encoding part of the Transformer architecture for reconstructing information on the cosmic particle mass. Exemplified are nine sensors (circles, central sensor in blue) with their information coded in $130$ latent variables (gray boxes). The red part shows the trainable Positional Encoding vectors of equal dimension which are added to the latent variables.}
\label{fig:Positional_Encoding}
\end{figure}

Figure~\ref{fig:Positional_Encoding} illustrates for a subset of nine sensors the positional encoding step within the complete Transformer architecture, developed for the conditions of the Pierre Auger Observatory \cite{PierreAuger:2023tyq}. Note that the two-dimensional structure of the sensor array is flattened into a one-dimensional vector, i.e. it is to be re-learned by trainable positional encoding. Prior to this encoding step, the sensor signals in the time domain have already been analysed, so that the most relevant information for each sensor is distilled into $d=130$ latent variables. A trainable positional encoding of the same dimension $d$ is then added to each sensor’s set of $d$ latent variables, preserving the dimensionality for the subsequent network stage that ultimately infers the mass-related parameters. For additional details on the network architecture, refer to \cite{PierreAuger:2023tyq}.

Here, we interpret the positional encodings for each sensor that were adapted during the network’s training to retrieve information on the particle mass. Typically, in an air shower event, one of the sensors exhibits the strongest signal, which we take then as the geometric center of the sensor array. The sensors in the immediately neighboring hexagonal ring record smaller signals, and sensors in successive outer rings are struck even less.

Our primary interest lies in the similarity of the $d=130$-dimensional positional encoding vectors across sensors. As a measure of similarity, we calculate the scalar product normalized to unity (similar to the cosine of the angle between vectors) by comparing the positional encoding vector $\vec{n}$ of one sensor to the positional encoding vectors $\vec{m}$ of every other sensor:
\begin{align}
    (\cos{\vartheta})_{\vec{n}\,\vec{m}}=\frac{\sum_{\ell=1}^{\ell=d} m_\ell\,n_\ell}{\vert\vec{n}\vert\,\vert\vec{m}\vert}
    \label{eq:scalar_product}
\end{align}
Values close to $\cos=1$ indicate a high similarity in positional encoding.

\begin{figure}[h]
\centering
\includegraphics[width=0.45\textwidth]{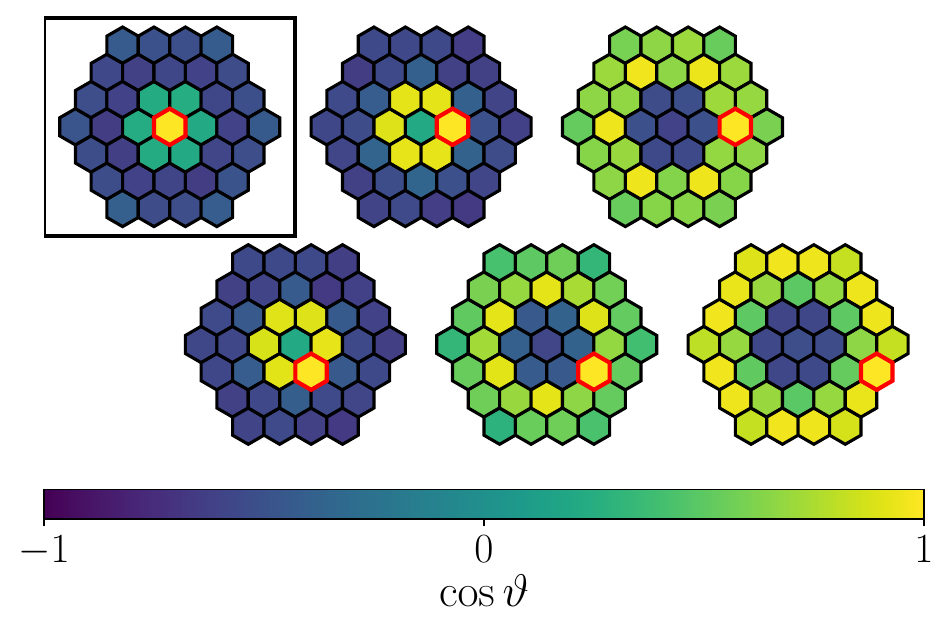}
\caption{The similarity measure of the positional encodings (\ref{eq:scalar_product}) between one highlighted sensor and its neighbors reveals the rotational symmetry of the hexagonal sensor structure which is learned by the Transformer.}
\label{fig:Hexagon_Symmetry}
\end{figure}

In the boxed upper left of Fig.~\ref{fig:Hexagon_Symmetry}, the sensor with the encoding vector $\vec{n}$ is marked by a red hexagon. In this case it refers to the sensor with the overall largest recorded signal. The inner color scale encodes the normalized scalar product (\ref{eq:scalar_product}). All sensors in the next hexagonal ring share roughly the same similarity value (shown in green), illustrating the rotational symmetry of the air shower. Sensors farther outward show smaller similarity values (in blue) relative to the most strongly hit sensor at the center.

The remaining five sub-figures show corresponding normalized scalar products (\ref{eq:scalar_product}) using as positional encoding vectors $\vec{n}$ those of adjacent sensors. The second sub-figure in the first row compares each sensor to the sensor adjacent to the right of the most strongly hit sensor. Once again, the color scale reveals a hexagonal rotational symmetry in yellow, i.e. the six positional encoding vectors are very similar with $\cos{\vartheta}\approx 1$. 

Further right, the subsequent sensor serves as the reference for computing scalar products (\ref{eq:scalar_product}), and remarkably, the hexagonal rotational symmetry emerges again—skipping one sensor neighbor to complete the ring of six similar positional encodings (again in yellow). The sub-figures below show normalized scalar products (\ref{eq:scalar_product}) with positional encoding vectors $\vec{n}$ of sensors in the row underneath the overall largest recorded signal, which leads to similar conclusions. We show only a fraction of possible sub-figures here, as the same patterns are also learned in all other directions. 

Note again that the network architecture itself contains no explicit information about symmetries. The network learns purely from the simulated air shower data that these events exhibit azimuthal rotational symmetry, which it encodes to improve reconstruction accuracy of the relevant mass-related observables. Although it is not shown here explicity, the potential to learn corrections such as Earth magnetic field effects can naturally be incorporated in the learned positional encodings.

\section{Attention}

\paragraph{Scientific context}
With successful detection of ultra-high-energy cosmic particles, the ensuing question is which galaxies could have accelerated them to such extreme energies. This is however complicated, as astronomical observations indicate that our Galaxy possesses a magnetic field capable of significantly deflecting ionized atomic nuclei. Although detailed magnetic field models exist \cite{Pshirkov:2011um,Jansson:2012rt,Unger:2023lob}, they are subject to substantial uncertainties \cite{Unger:2017kfh}, making it challenging to pinpoint the sources of cosmic particles based on their observed arrival directions. 

The guiding scientific question of this study is thus: given a galaxy catalog of source candidates, can one demonstrate by modifying the deflection angles and strengths of a galactic magnetic field model that at least a fraction of the observed cosmic particles originate from these cataloged galaxies. If successful, cosmic ray sources would be identified together with evidence of magnetic field deflections \cite{Schulte:2023iry}. 

\begin{figure}[h]
\centering
\includegraphics[width=0.31\textwidth]{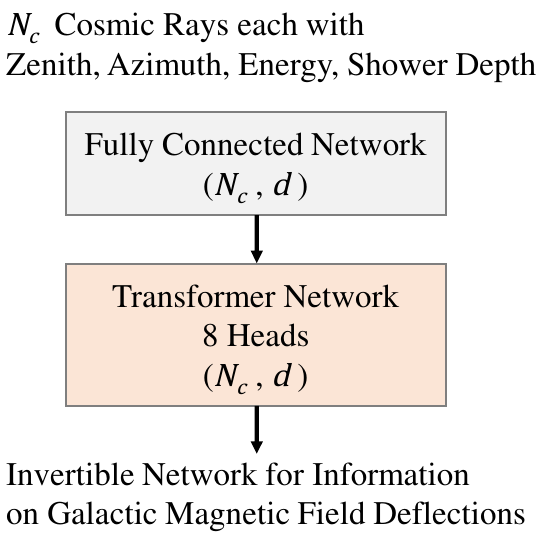}
\caption{Transformer part of the network architecture to adapt a galactic magnetic field model according to cosmic rays from a galaxy catalog. In total $8$ Transformer heads separate signal particles originating from these galaxies and background particles from elsewhere.}
\label{fig:Classification_Network}
\end{figure}

Here, the Transformer network serves as a preprocessing tool for the observed cosmic particles, allowing an invertible network to adapt the directions and magnitudes of Galactic deflections via harmonic coefficients (Fig.~\ref{fig:Classification_Network}).

We trained both the invertible network and the Transformer jointly on $\mathcal{O}(10^6)$ astrophysical simulations \cite{AlvesBatista:2022vem}, with each simulation containing on the order of $N_{c} \approx 4{,}000$ cosmic particles. In these simulations, $10\%$ of the particles are designated as signal particles accelerated in galaxies from a catalog of $\gamma$-AGNs \cite{PierreAuger:2023htc}, while $90\%$ are injected additionally as background particles. The signal particles are deflected differently in each simulation through variations of the harmonic coefficients to train the invertible network. 

\begin{figure}[h]
\centering
\includegraphics[width=0.5\textwidth]{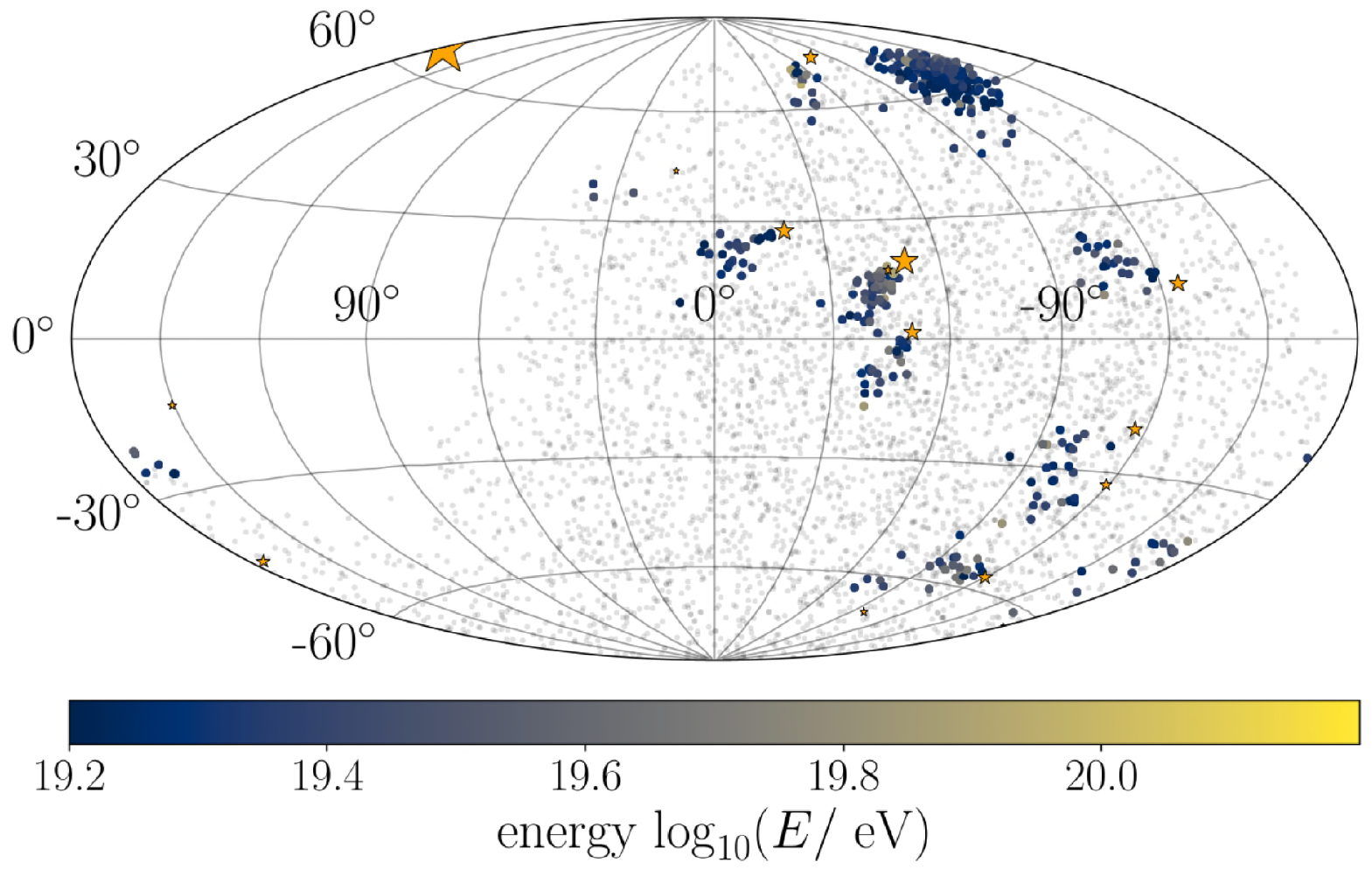}
\caption{Example of a simulated astrophysical scenario in galactic coordinates with the star symbols representing the cosmic-ray origins, the colored symbols denote signal cosmic rays from the galaxies which were deflected in the galactic magnetic field, and gray symbols are background particles.}
\label{fig:CR_Simulated_Dataset}
\end{figure}

Figure~\ref{fig:CR_Simulated_Dataset} shows one of the astrophysical scenarios in galactic coordinates where star symbols denote the source directions, signal particles from the sources are color coded with their energy, and gray symbols denote background particles. Signal particles exemplify local coherent deflections in the galactic magnetic field. The density of cosmic rays follows the observatory view of the Pierre Auger Observatory.

\paragraph{Research on Transformer}
The Transformer’s primary task is to determine which cosmic particles are candidates for originating from the catalog’s galaxies (signal) and which presumably come from elsewhere (background). Ideally, adjustments to the galactic magnetic field should be driven only by the signal particles. To understand why this selection works quite effectively, we analysed the Transformer’s attention values.

Because of the large number of cosmic particles, the memory requirements for a standard Transformer exceeded our GPU capacity (NVIDIA RTX 6000 Ada Generation). Consequently, we employed the Nyströmformer \cite{xiong2021}, a specialized Transformer variant in which internal dimensions are deliberately reduced, though an approximate attention matrix can be recovered with the same shape as in a conventional Transformer.

For each astrophysical scenario $i$ we analyse the attentions assigned to the $N_{c}$ cosmic particles by the scaled-dot-product attention \cite{Vaswani:2017lxt}:
\begin{align}
    \text{softmax}\Big(\frac{Q\,K^T}{\sqrt{d^\prime}} \Big)\,V
    \label{eq:attention-QKV}
\end{align}
In the following we explain the path to visualizing self-attentions in a sky map starting from equation (\ref{eq:attention-QKV}).
The matrix $Q$ referred to as \textit{query} is calculated from a trainable weight matrix of dimensions $d\times d^\prime$ applied to the $N_c$ cosmic particles with their $d$ latent variables:
\begin{align}
Q&=
{\tabrotate{\footnotesize$N_{c}$}} \,
\overset{\text{Latent}}{\underset{d}{
\begin{NiceTabular}{*{3}{c}}[hvlines,cell-space-top-limit=3pt]
&&\\
\\
\end{NiceTabular}}}\hspace*{3mm}
{\tabrotate{\footnotesize $d$}} \,
\overset{Q-\text{Weights}}{\underset{d^\prime}{
\begin{NiceTabular}{*{4}{c}}%
[hvlines,cell-space-top-limit=3pt]
&&&\\
\\
\\
\end{NiceTabular}}} \nonumber\\
&={\tabrotate{\footnotesize$N_{c}$}} \,
\underset{d^\prime}{
\begin{NiceTabular}{*{4}{c}}[hvlines,cell-space-top-limit=3pt]
&&&\\
\\
\end{NiceTabular}}
\label{eq:QKV-matrices}
\end{align}
Both the matrices $K$ (\textit{key}) and $V$ (\textit{value}) are calculated accordingly with their own trainable weight matrices.

The matrix product $Q$ times the transposed of $K$ in (\ref{eq:attention-QKV}) gives a matrix of dimension $N_c\times N_c$:
\begin{align}
Q\,K^T&={\tabrotate{\footnotesize $N_{c}$}} \,
\underset{d^\prime}{
\begin{NiceTabular}{*{4}{c}}[hvlines,cell-space-top-limit=3pt]
&&&\\
\\
\end{NiceTabular}}\hspace*{3mm}
{\tabrotate{\footnotesize $d^\prime$}} \,
\underset{N_{c}}{
\begin{NiceTabular}{*{2}{c}}[hvlines,cell-space-top-limit=3pt]
&\\
\\
\\
\\
\end{NiceTabular}}\nonumber\\
&={\tabrotate{\footnotesize $N_{c}$}} \,
\underset{N_{c}}{
\begin{NiceTabular}{*{2}{c}}[hvlines,cell-space-top-limit=3pt]
&\\
\\
\end{NiceTabular}}
\end{align}
Normalization by the square root of the \textit{query} and \textit{key} dimension $d^\prime$ and applying the softmax-function to each row (indicated by the gray color) results in the attention weights, preserving the $N_c\times N_c$ shape:
\begin{align}
\text{softmax}\Big(\frac{Q\,K^T}{\sqrt{d^\prime}}\Big)&=
{\tabrotate{\footnotesize $N_{c}$}} \,
\underset{N_{c}}{
\begin{NiceArray}{llll}[hvlines, code-before = \cellcolor{gray!15}{1-1,1-2}]
\hspace*{3mm} &  \hspace*{3mm} \\
\hspace*{3mm} &  \hspace*{3mm} \\
\end{NiceArray}}
\label{eq:attention-weight-matrix}
\end{align}
We also show how the calculation of equation (\ref{eq:attention-QKV}) is completed for subsequently explaining the visualization of the cosmic-particle attentions (\ref{eq:attention-weight-matrix}).
The attention weight matrix is multiplied with the \textit{value} matrix $V$:
\begin{align}
A(Q,K,V)&=\text{softmax}\Big(\frac{Q\,K^T}{\sqrt{d^\prime}}\Big)\,V\nonumber\\
&=
{\tabrotate{\footnotesize $N_{c}$}} \,
\underset{N_{c}}{
\begin{NiceArray}{ll}[hvlines, code-before = \cellcolor{red!15}{1-1,2-1}\cellcolor{blue!15}{1-2,2-2}]
\hspace*{3mm} &  \hspace*{3mm} \\
 &  \\
\end{NiceArray}}\hspace*{3mm}
{\tabrotate{\footnotesize$N_{c}$}} \,
\underset{d^\prime}{
\begin{NiceArray}{llll}[hvlines, code-before = \cellcolor{red!15}{1-1,1-2,1-3,1-4}\cellcolor{blue!15}{2-1,2-2,2-3,2-4}]
\hspace*{3mm} &  \hspace*{3mm} & \hspace*{3mm} &  \hspace*{3mm} \\
 &  \\
\end{NiceArray}}\\
&={\tabrotate{\footnotesize$N_{c}$}} \,
\underset{d^\prime}{\begin{NiceTabular}{*{4}{c}}[hvlines,cell-space-top-limit=3pt]
&&&\\
\\
\end{NiceTabular}}
\end{align}
Remember, the \textit{value} matrix $V$ results from the same calculation explained in (\ref{eq:QKV-matrices}) by replacing $Q$ with $V$. Each row contains the features of one cosmic particle, as indicated by the red and blue colors. For obtaining the output $A$, the \textit{columns of the attention matrix} (\ref{eq:attention-weight-matrix}) provide the weights for the individual cosmic-particle latent features during matrix multiplication.

Thus, we sum the columns of the attention matrix and subsequently normalize all summed attentions to one in order to obtain for every cosmic particle a relative attention value:
\begin{align}
\vec{A}^\prime(N_C)&=\frac{1}{\text{Norm}}\,\Big(\sum_{\text{column}} \hspace*{1mm}
{\tabrotate{\footnotesize$N_{c}$}} \,
\underset{N_{c}}{
\begin{NiceArray}{ll}[hvlines, code-before = \cellcolor{red!15}{1-1,2-1}\cellcolor{blue!15}{1-2,2-2}]
\hspace*{3mm} &  \hspace*{3mm} \\
 & \\
\end{NiceArray}}\;\Big)\nonumber\\
&=
\underset{N_{c}}{
\begin{NiceArray}{ll}[hvlines, code-before = \cellcolor{red!15}{1-1}\cellcolor{blue!15}{1-2}]
\hspace*{3mm} &  \hspace*{3mm} \\
\end{NiceArray}}
\end{align}

Then we transform to the healpix coordinate system with $n_{side}=8$ corresponding to $M=768$ pixels \cite{Gorski:2004by,Zonca:2019vzt}. We add up the normalized attentions of particles contributing to a pixel and normalize to the number of contributing particles:
\begin{align}
\vec{A}(\text{Pixel})&=\frac{1}{\text{Norm}_{\text{pix}}}\hspace*{3mm}
\underset{N_{\text{pix}}}{
\begin{NiceArray}{ll}[hvlines]
\hspace*{3mm} &  \hspace*{3mm} \\
\end{NiceArray}}
\end{align}
In this way we get a sky map of the relative attentions $\vec{A}_i(\text{Pixel})$ for the astrophysical scenario $i$.

\begin{figure}[h]
\centering
\includegraphics[width=0.4\textwidth]{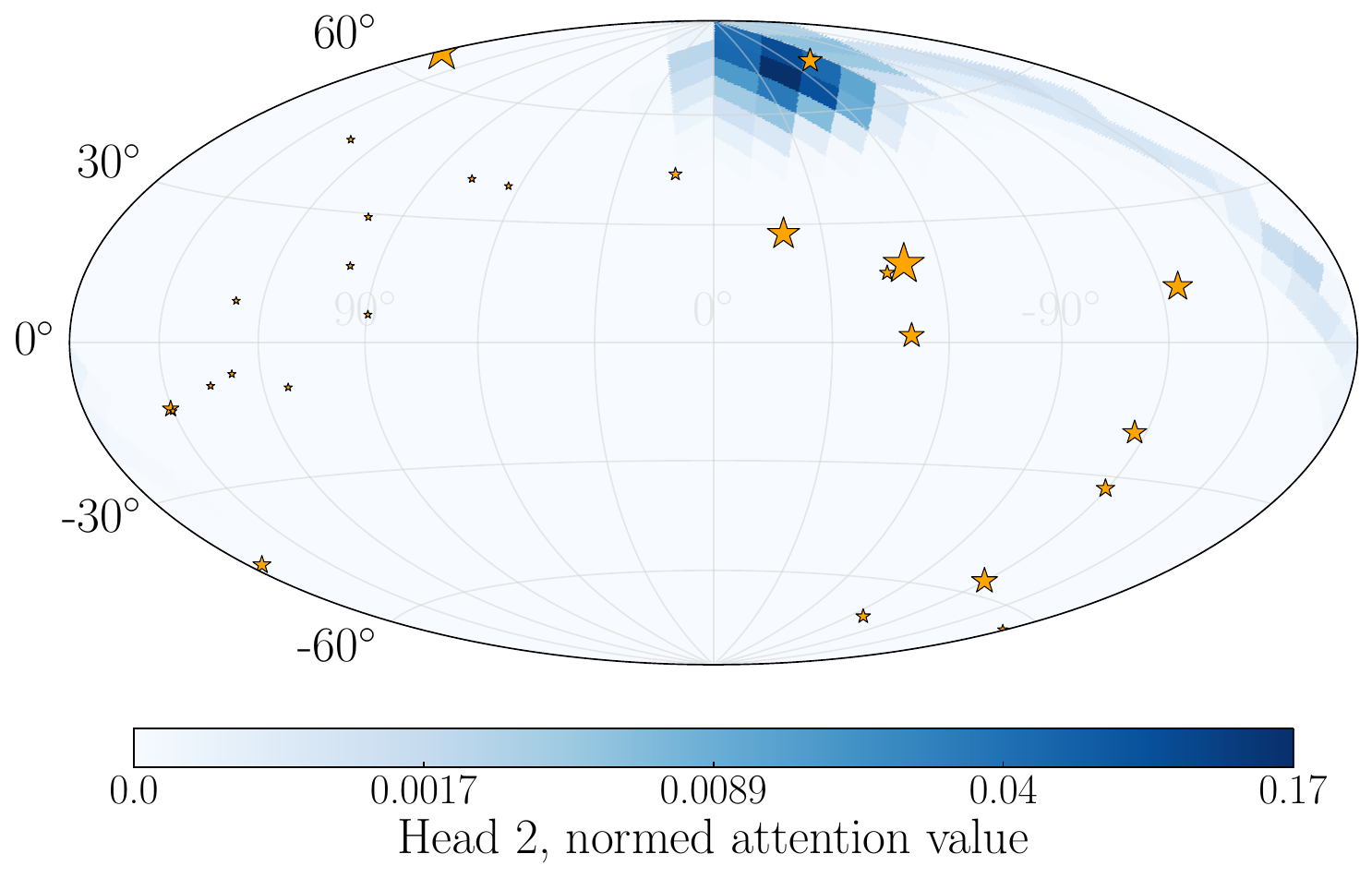}
\includegraphics[width=0.4\textwidth]{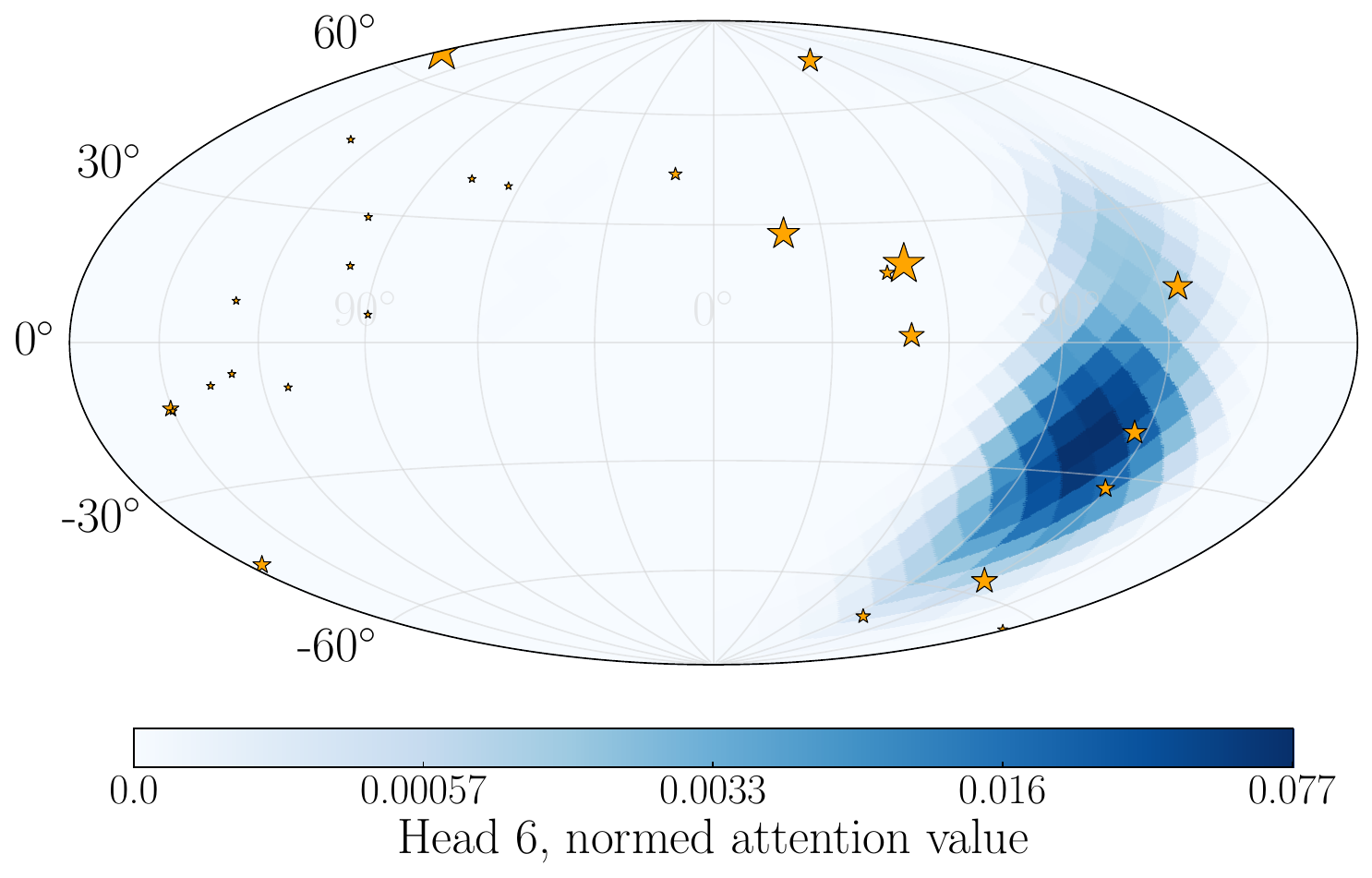}
\includegraphics[width=0.4\textwidth]{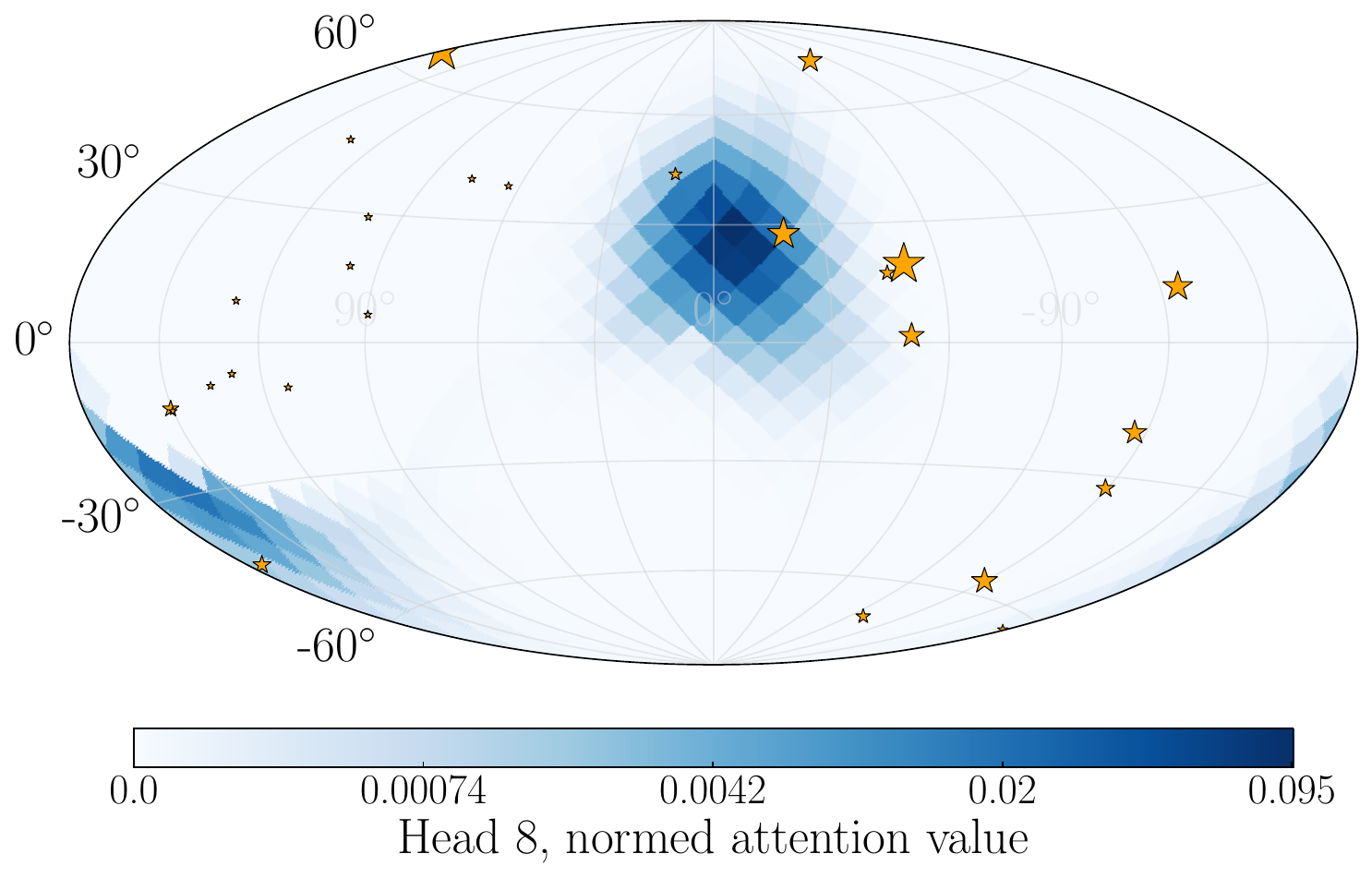}
\caption{Normalized attention sky maps in galactic coordinates resulting from $1,000$ astrophysical scenarios, exemplarily for three of the eight Transformer heads. The color scale is logarithmic indicating regions of high attention values by light colors.}
\label{fig:Attention_sky}
\end{figure}

Figure~\ref{fig:Attention_sky} shows the attention average of $1,000$ astrophysical scenarios exemplarily for three of the Transformer heads. Remember that the origins of the astrophysical scenarios are fixed to the galaxy catalog, but the galactic magnetic field model is modified by varying the spherical harmonic coefficients. The color scale was chosen to be logarithmic in order to visualize also small attention values. High attention means that the Transformer has been trained searching for signal particles preferred in these regions. Maximum attention is slightly shifted from the galaxy origins as the galactic field model requires coherent deflections whose direction and strength are modified by the harmonic coefficients.

It can well be seen that each of the heads concentrates on a specific region in the sky trying to identify signal particle candidates. This is true also for the remaining heads that are not shown here. 

\begin{figure}[h]
\centering
\includegraphics[width=0.4\textwidth]{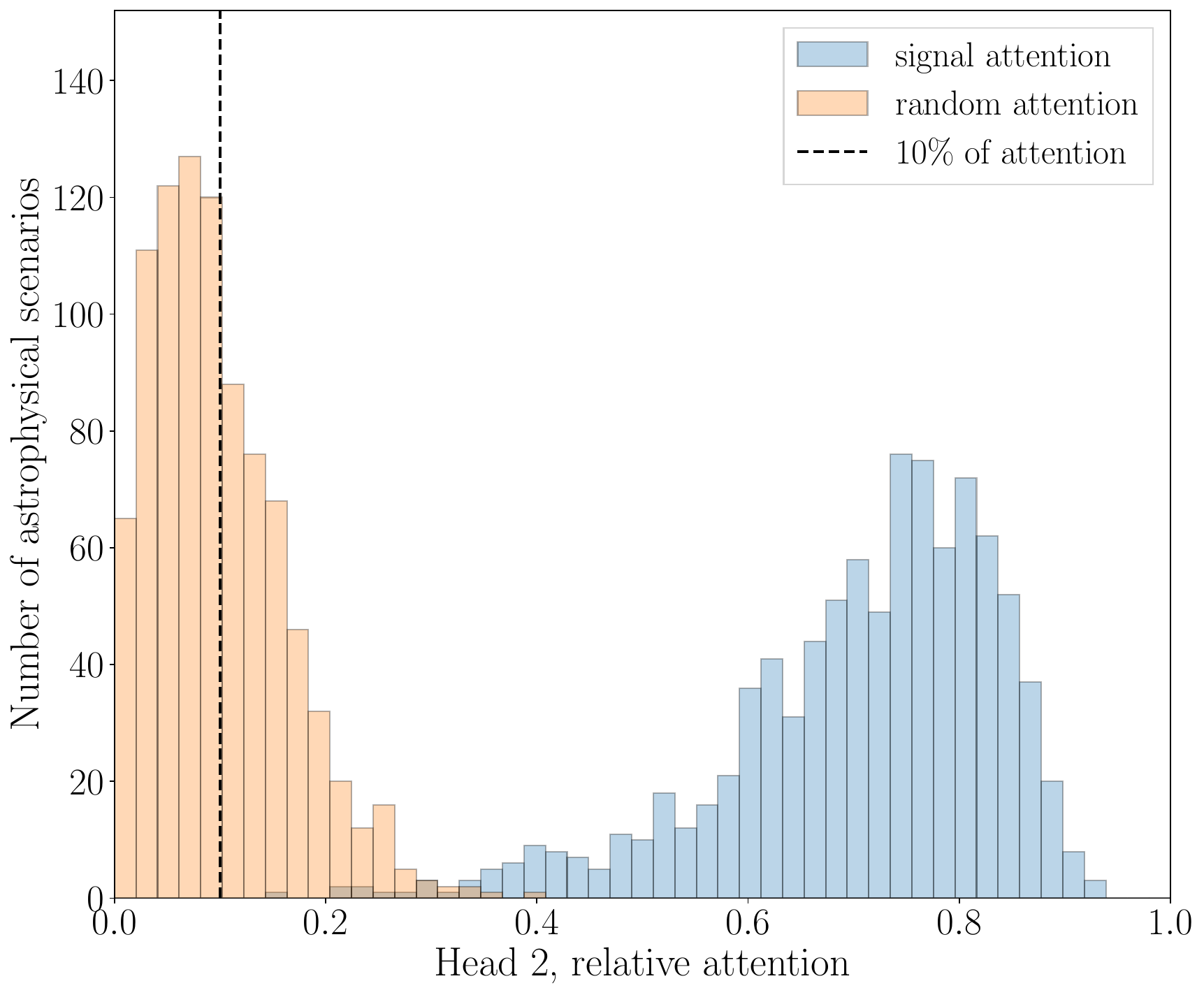}
\includegraphics[width=0.4\textwidth]{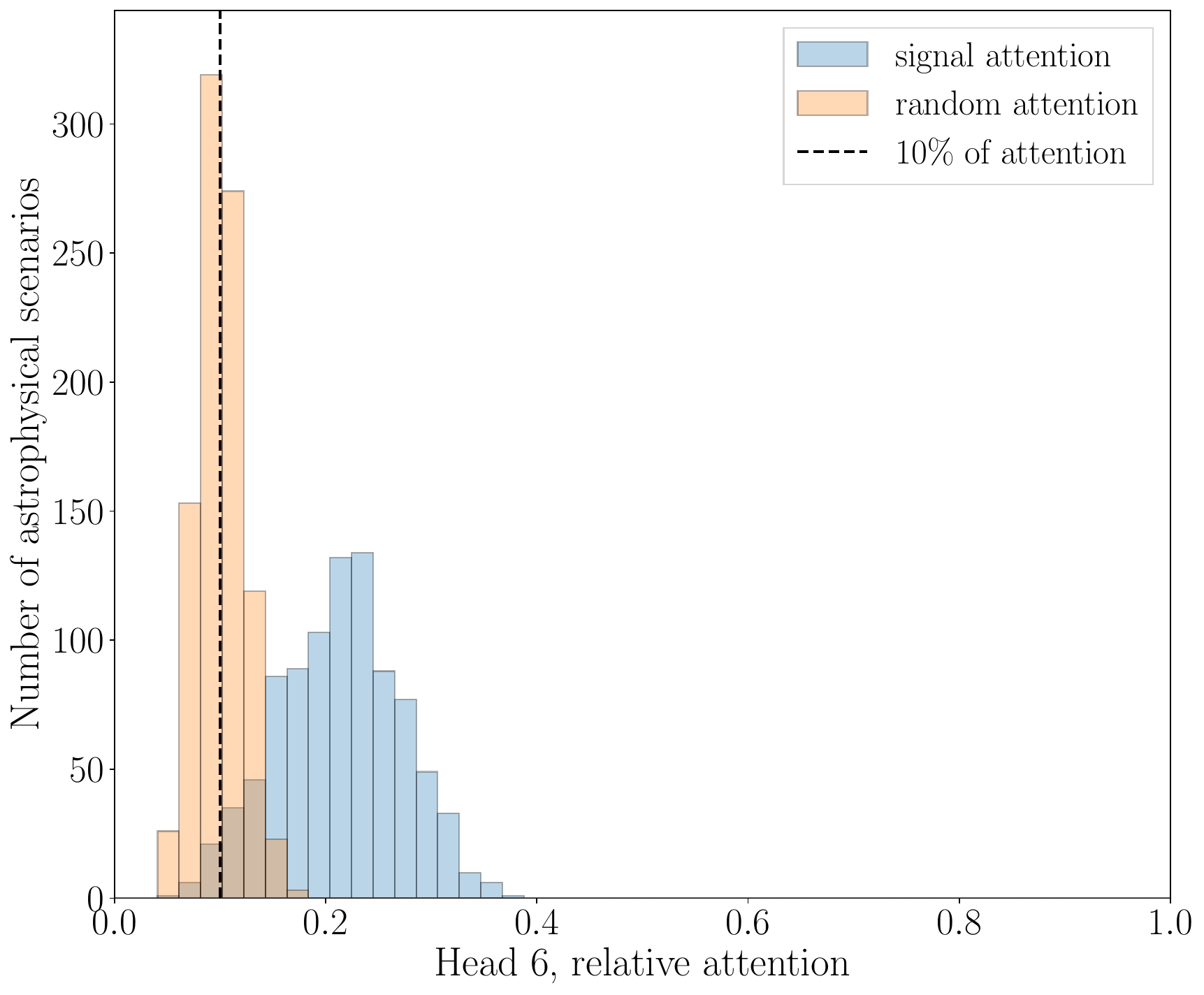}
\includegraphics[width=0.4\textwidth]{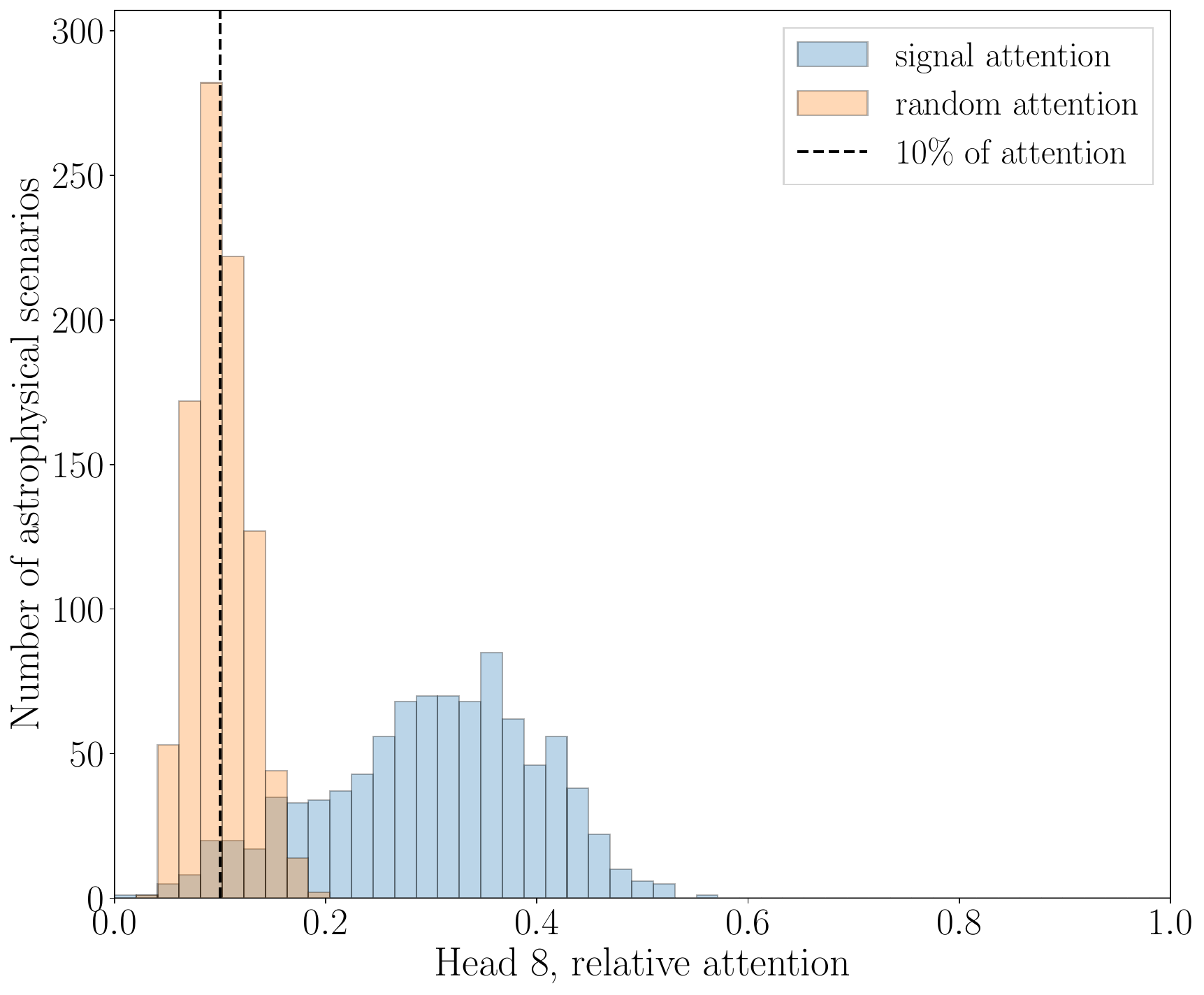}
\caption{Sum of the normalized attention values of the signal cosmic rays for $1,000$ astrophysical scenarios as blue histogram. The background attention is shown by the orange histogram.}
\label{fig:Attention_separation}
\end{figure}

Below, we investigate the rate at which the Transformer identifies signal particles in a previously unknown astrophysical scenarios. For a new astrophysical scenario $i$, we sum the attention values $a_i$ corresponding to the signal particles, which amount to $10\%$ of the total number of cosmic particles and whose identity we know from the simulation. As a benchmark, we add up the attention values of an equally sized ($10\%$) random selection of background particles. In the top Figure~\ref{fig:Attention_separation}, for $1,000$ scenarios, the aggregated attention for signal particles is shown as a blue histogram, and that for background particles as an orange histogram, focusing on Transformer head $2$. The attention values of signal cosmic rays are well separated from background cosmic rays. The additional histograms of Figure~\ref{fig:Attention_separation} present results for Transformer heads $6$ and $8$, which exhibit slightly lower discrimination power yet still show clear separations. Thus, the Transformer assigns large attention values to signal cosmic particles originating from the given galaxy catalog.

Finally, we examined the importance of the input variables — energy, arrival direction (zenith and azimuth), and mass information through shower depth — following an integrated gradients approach as described in \cite{sundararajan2017}. Our findings indicate that directional information dominates within the attention mechanism. In addition, both energy and to a lesser extent also shower depth contribute to the attention.

In summary of this section, the attention mechanism focuses on the sky directions and the necessary properties of the cosmic rays that would make them candidates for originating from the given galaxy catalog. Note that in the case of an isotropic sky, the simple transformer used here would also search for cosmic rays carrying the properties trained from the astrophysical scenario with the galaxy catalog.

\section{Conclusion}
\label{sec:Conclusion}

We examined Transformer networks in two distinct ultra-high-energy cosmic ray applications. First, the Transformer learned to take advantage of azimuthal symmetry from simulated air shower data with a hexagonal sensor arrangement. We found that this information was trained into the Transformer’s positional encodings. Second, we tackled understanding a classification challenge for cosmic particles originating from specific galaxies or from background. We found that each Transformer head focuses on a different region in the sky and marks particles originating from the galaxy with higher attention values than background particles.

We thus exemplified two transformer applications where we obtained deeper insight into the work performance of trainable positional encodings and the attention mechanism.

\backmatter

\bmhead{Supplementary information}

Not applicable

\bmhead{Acknowledgements}

This work is supported by the Ministry of Innovation, Science, and Research of the State of North Rhine-Westphalia, and by the Federal Ministry of Education and Research (BMBF) in Germany. Language translation support was received by deepL and Open AI o1.

\section*{Declarations}

The authors have no competing interests to declare that are relevant to the content of this article.

%Some journals require declarations to be submitted in a standardised format. Please check the Instructions for Authors of the journal to which you are submitting to see if you need to complete this section. If yes, your manuscript must contain the following sections under the heading `Declarations':

%\begin{itemize}
%\item Conflict of interest/Competing interests (check journal-specific guidelines for which heading to use)
%\item Ethics approval and consent to participate
%\item Consent for publication
%\item Data availability 
%\end{itemize}

%\noindent
%If any of the sections are not relevant to your manuscript, please include the heading and write %`Not applicable' for that section. 

%\bibliography{attention}% common bib file


\begin{thebibliography}{26}
\ifx \bisbn   \undefined \def \bisbn  #1{ISBN #1}\fi
\ifx \binits  \undefined \def \binits#1{#1}\fi
\ifx \bauthor  \undefined \def \bauthor#1{#1}\fi
\ifx \batitle  \undefined \def \batitle#1{#1}\fi
\ifx \bjtitle  \undefined \def \bjtitle#1{#1}\fi
\ifx \bvolume  \undefined \def \bvolume#1{\textbf{#1}}\fi
\ifx \byear  \undefined \def \byear#1{#1}\fi
\ifx \bissue  \undefined \def \bissue#1{#1}\fi
\ifx \bfpage  \undefined \def \bfpage#1{#1}\fi
\ifx \blpage  \undefined \def \blpage #1{#1}\fi
\ifx \burl  \undefined \def \burl#1{\textsf{#1}}\fi
\ifx \doiurl  \undefined \def \doiurl#1{\url{https://doi.org/#1}}\fi
\ifx \betal  \undefined \def \betal{\textit{et al.}}\fi
\ifx \binstitute  \undefined \def \binstitute#1{#1}\fi
\ifx \binstitutionaled  \undefined \def \binstitutionaled#1{#1}\fi
\ifx \bctitle  \undefined \def \bctitle#1{#1}\fi
\ifx \beditor  \undefined \def \beditor#1{#1}\fi
\ifx \bpublisher  \undefined \def \bpublisher#1{#1}\fi
\ifx \bbtitle  \undefined \def \bbtitle#1{#1}\fi
\ifx \bedition  \undefined \def \bedition#1{#1}\fi
\ifx \bseriesno  \undefined \def \bseriesno#1{#1}\fi
\ifx \blocation  \undefined \def \blocation#1{#1}\fi
\ifx \bsertitle  \undefined \def \bsertitle#1{#1}\fi
\ifx \bsnm \undefined \def \bsnm#1{#1}\fi
\ifx \bsuffix \undefined \def \bsuffix#1{#1}\fi
\ifx \bparticle \undefined \def \bparticle#1{#1}\fi
\ifx \barticle \undefined \def \barticle#1{#1}\fi
\bibcommenthead
\ifx \bconfdate \undefined \def \bconfdate #1{#1}\fi
\ifx \botherref \undefined \def \botherref #1{#1}\fi
\ifx \url \undefined \def \url#1{\textsf{#1}}\fi
\ifx \bchapter \undefined \def \bchapter#1{#1}\fi
\ifx \bbook \undefined \def \bbook#1{#1}\fi
\ifx \bcomment \undefined \def \bcomment#1{#1}\fi
\ifx \oauthor \undefined \def \oauthor#1{#1}\fi
\ifx \citeauthoryear \undefined \def \citeauthoryear#1{#1}\fi
\ifx \endbibitem  \undefined \def \endbibitem {}\fi
\ifx \bconflocation  \undefined \def \bconflocation#1{#1}\fi
\ifx \arxivurl  \undefined \def \arxivurl#1{\textsf{#1}}\fi
\csname PreBibitemsHook\endcsname

%%% 1
\bibitem[\protect\citeauthoryear{Vaswani et~al.}{2017}]{Vaswani:2017lxt}
\begin{botherref}
\oauthor{\bsnm{Vaswani}, \binits{A.}},
\oauthor{\bsnm{Shazeer}, \binits{N.}},
\oauthor{\bsnm{Parmar}, \binits{N.}},
\oauthor{\bsnm{Uszkoreit}, \binits{J.}},
\oauthor{\bsnm{Jones}, \binits{L.}},
\oauthor{\bsnm{Gomez}, \binits{A.N.}},
\oauthor{\bsnm{Kaiser}, \binits{L.}},
\oauthor{\bsnm{Polosukhin}, \binits{I.}}:
{Attention Is All You Need}
(2017)
{\href{https://arxiv.org/abs/1706.03762}{{arXiv:1706.03762}}}
{[cs.CL]}
\end{botherref}
\endbibitem

%%% 2
\bibitem[\protect\citeauthoryear{Dosovitskiy et~al.}{2020}]{Dosovitskiy:2020qjv}
\begin{botherref}
\oauthor{\bsnm{Dosovitskiy}, \binits{A.}}, et al.:
{An Image is Worth 16x16 Words: Transformers for Image Recognition at Scale}
(2020)
{\href{https://arxiv.org/abs/2010.11929}{{arXiv:2010.11929}}}
{[cs.CV]}
\end{botherref}
\endbibitem

%%% 3
\bibitem[\protect\citeauthoryear{OpenAI}{2022}]{openai_chatgpt_2022}
\begin{botherref}
\oauthor{\bsnm{OpenAI}}:
ChatGPT: Optimizing Language Models for Dialogue.
\url{https://openai.com/blog/chatgpt/}.
Online; Zugriff am 30. November 2022
(2022)
\end{botherref}
\endbibitem

%%% 4
\bibitem[\protect\citeauthoryear{Qu et~al.}{2022}]{Qu:2022mxj}
\begin{botherref}
\oauthor{\bsnm{Qu}, \binits{H.}},
\oauthor{\bsnm{Li}, \binits{C.}},
\oauthor{\bsnm{Qian}, \binits{S.}}:
{Particle Transformer for Jet Tagging}
(2022)
{\href{https://arxiv.org/abs/2202.03772}{{arXiv:2202.03772}}}
{[hep-ph]}
\end{botherref}
\endbibitem

%%% 5
\bibitem[\protect\citeauthoryear{Finke et~al.}{2023}]{Finke:2023veq}
\begin{barticle}
\bauthor{\bsnm{Finke}, \binits{T.}},
\bauthor{\bsnm{Kr\"amer}, \binits{M.}},
\bauthor{\bsnm{M\"uck}, \binits{A.}},
\bauthor{\bsnm{T\"onshoff}, \binits{J.}}:
\batitle{{Learning the language of QCD jets with transformers}}.
\bjtitle{JHEP}
\bvolume{06},
\bfpage{184}
(\byear{2023})
\doiurl{10.1007/JHEP06(2023)184}
{\href{https://arxiv.org/abs/2303.07364}{{arXiv:2303.07364}}}
{[hep-ph]}
\end{barticle}
\endbibitem

%%% 6
\bibitem[\protect\citeauthoryear{Abdul~Halim et~al.}{2023}]{PierreAuger:2023tyq}
\begin{barticle}
\bauthor{\bsnm{Abdul~Halim}, \binits{A.}}, \betal:
\batitle{{Deep-Learning-Based Cosmic-Ray Mass Reconstruction Using the Water-Cherenkov and Scintillation Detectors of AugerPrime}}.
\bjtitle{PoS}
\bvolume{ICRC2023},
\bfpage{371}
(\byear{2023})
\doiurl{10.22323/1.444.0371}
\end{barticle}
\endbibitem

%%% 7
\bibitem[\protect\citeauthoryear{Wu et~al.}{2025}]{Wu:2024thh}
\begin{barticle}
\bauthor{\bsnm{Wu}, \binits{Y.}},
\bauthor{\bsnm{Wang}, \binits{K.}},
\bauthor{\bsnm{Li}, \binits{C.}},
\bauthor{\bsnm{Qu}, \binits{H.}},
\bauthor{\bsnm{Zhu}, \binits{J.}}:
\batitle{{Jet tagging with more-interaction particle transformer}}.
\bjtitle{Chin. Phys. C}
\bvolume{49}(\bissue{1}),
\bfpage{013110}
(\byear{2025})
\doiurl{10.1088/1674-1137/ad7f3d}
{\href{https://arxiv.org/abs/2407.08682}{{arXiv:2407.08682}}}
{[hep-ph]}
\end{barticle}
\endbibitem

%%% 8
\bibitem[\protect\citeauthoryear{Brehmer et~al.}{2024}]{Brehmer:2024yqw}
\begin{botherref}
\oauthor{\bsnm{Brehmer}, \binits{J.}},
\oauthor{\bsnm{Bres\'o}, \binits{V.}},
\oauthor{\bsnm{Haan}, \binits{P.}},
\oauthor{\bsnm{Plehn}, \binits{T.}},
\oauthor{\bsnm{Qu}, \binits{H.}},
\oauthor{\bsnm{Spinner}, \binits{J.}},
\oauthor{\bsnm{Thaler}, \binits{J.}}:
{A Lorentz-Equivariant Transformer for All of the LHC}
(2024)
{\href{https://arxiv.org/abs/2411.00446}{{arXiv:2411.00446}}}
{[hep-ph]}
\end{botherref}
\endbibitem

%%% 9
\bibitem[\protect\citeauthoryear{Feickert and Nachman}{2021}]{Feickert:2021ajf}
\begin{botherref}
\oauthor{\bsnm{Feickert}, \binits{M.}},
\oauthor{\bsnm{Nachman}, \binits{B.}}:
{A Living Review of Machine Learning for Particle Physics}
(2021)
{\href{https://arxiv.org/abs/2102.02770}{{arXiv:2102.02770}}}
{[hep-ph]}
\end{botherref}
\endbibitem

%%% 10
\bibitem[\protect\citeauthoryear{Wang et~al.}{2024}]{Wang:2024rup}
\begin{botherref}
\oauthor{\bsnm{Wang}, \binits{A.}},
\oauthor{\bsnm{Gandrakota}, \binits{A.}},
\oauthor{\bsnm{Ngadiuba}, \binits{J.}},
\oauthor{\bsnm{Sahu}, \binits{V.}},
\oauthor{\bsnm{Bhatnagar}, \binits{P.}},
\oauthor{\bsnm{Khoda}, \binits{E.E.}},
\oauthor{\bsnm{Duarte}, \binits{J.}}:
{Interpreting Transformers for Jet Tagging}
(2024)
{\href{https://arxiv.org/abs/2412.03673}{{arXiv:2412.03673}}}
{[hep-ph]}
\end{botherref}
\endbibitem

%%% 11
\bibitem[\protect\citeauthoryear{Aab et~al.}{2015}]{PierreAuger:2015eyc}
\begin{barticle}
\bauthor{\bsnm{Aab}, \binits{A.}}, \betal:
\batitle{{The Pierre Auger Cosmic Ray Observatory}}.
\bjtitle{Nucl. Instrum. Meth. A}
\bvolume{798},
\bfpage{172}--\blpage{213}
(\byear{2015})
\doiurl{10.1016/j.nima.2015.06.058}
{\href{https://arxiv.org/abs/1502.01323}{{arXiv:1502.01323}}}
{[astro-ph.IM]}
\end{barticle}
\endbibitem

%%% 12
\bibitem[\protect\citeauthoryear{Aab et~al.}{2021a}]{PierreAuger:2021fkf}
\begin{barticle}
\bauthor{\bsnm{Aab}, \binits{A.}}, \betal:
\batitle{{Deep-learning based reconstruction of the shower maximum $X_{max}$ using the water-Cherenkov detectors of the Pierre Auger Observatory}}.
\bjtitle{JINST}
\bvolume{16}(\bissue{07}),
\bfpage{07019}
(\byear{2021})
\doiurl{10.1088/1748-0221/16/07/P07019}
{\href{https://arxiv.org/abs/2101.02946}{{arXiv:2101.02946}}}
{[astro-ph.IM]}
\end{barticle}
\endbibitem

%%% 13
\bibitem[\protect\citeauthoryear{Aab et~al.}{2021b}]{PierreAuger:2021nsq}
\begin{barticle}
\bauthor{\bsnm{Aab}, \binits{A.}}, \betal:
\batitle{{Extraction of the muon signals recorded with the surface detector of the Pierre Auger Observatory using recurrent neural networks}}.
\bjtitle{JINST}
\bvolume{16}(\bissue{07}),
\bfpage{07016}
(\byear{2021})
\doiurl{10.1088/1748-0221/16/07/P07016}
{\href{https://arxiv.org/abs/2103.11983}{{arXiv:2103.11983}}}
{[hep-ex]}
\end{barticle}
\endbibitem

%%% 14
\bibitem[\protect\citeauthoryear{Abdul~Halim et~al.}{2025}]{PierreAuger:2024nzw}
\begin{barticle}
\bauthor{\bsnm{Abdul~Halim}, \binits{A.}}, \betal:
\batitle{{Measurement of the depth of maximum of air-shower profiles with energies between $10^{18.5}$ and $10^{20}$\,\,eV using the surface detector of the Pierre Auger Observatory and deep learning}}.
\bjtitle{Phys. Rev. D}
\bvolume{111}(\bissue{2}),
\bfpage{022003}
(\byear{2025})
\doiurl{10.1103/PhysRevD.111.022003}
{\href{https://arxiv.org/abs/2406.06319}{{arXiv:2406.06319}}}
{[astro-ph.HE]}
\end{barticle}
\endbibitem

%%% 15
\bibitem[\protect\citeauthoryear{Hoogeboom et~al.}{2018}]{Hoogeboom:2018exn}
\begin{botherref}
\oauthor{\bsnm{Hoogeboom}, \binits{E.}},
\oauthor{\bsnm{Peters}, \binits{J.W.T.}},
\oauthor{\bsnm{Cohen}, \binits{T.S.}},
\oauthor{\bsnm{Welling}, \binits{M.}}:
{HexaConv}
(2018)
{\href{https://arxiv.org/abs/1803.02108}{{arXiv:1803.02108}}}
{[cs.LG]}
\end{botherref}
\endbibitem

%%% 16
\bibitem[\protect\citeauthoryear{Pshirkov et~al.}{2011}]{Pshirkov:2011um}
\begin{barticle}
\bauthor{\bsnm{Pshirkov}, \binits{M.S.}},
\bauthor{\bsnm{Tinyakov}, \binits{P.G.}},
\bauthor{\bsnm{Kronberg}, \binits{P.P.}},
\bauthor{\bsnm{Newton-McGee}, \binits{K.J.}}:
\batitle{{Deriving global structure of the Galactic Magnetic Field from Faraday Rotation Measures of extragalactic sources}}.
\bjtitle{Astrophys. J.}
\bvolume{738},
\bfpage{192}
(\byear{2011})
\doiurl{10.1088/0004-637X/738/2/192}
{\href{https://arxiv.org/abs/1103.0814}{{arXiv:1103.0814}}}
{[astro-ph.GA]}
\end{barticle}
\endbibitem

%%% 17
\bibitem[\protect\citeauthoryear{Jansson and Farrar}{2012}]{Jansson:2012rt}
\begin{barticle}
\bauthor{\bsnm{Jansson}, \binits{R.}},
\bauthor{\bsnm{Farrar}, \binits{G.R.}}:
\batitle{{The Galactic Magnetic Field}}.
\bjtitle{Astrophys. J. Lett.}
\bvolume{761},
\bfpage{11}
(\byear{2012})
\doiurl{10.1088/2041-8205/761/1/L11}
{\href{https://arxiv.org/abs/1210.7820}{{arXiv:1210.7820}}}
{[astro-ph.GA]}
\end{barticle}
\endbibitem

%%% 18
\bibitem[\protect\citeauthoryear{Unger and Farrar}{2024}]{Unger:2023lob}
\begin{barticle}
\bauthor{\bsnm{Unger}, \binits{M.}},
\bauthor{\bsnm{Farrar}, \binits{G.R.}}:
\batitle{{The Coherent Magnetic Field of the Milky Way}}.
\bjtitle{Astrophys. J.}
\bvolume{970}(\bissue{1}),
\bfpage{95}
(\byear{2024})
\doiurl{10.3847/1538-4357/ad4a54}
{\href{https://arxiv.org/abs/2311.12120}{{arXiv:2311.12120}}}
{[astro-ph.GA]}
\end{barticle}
\endbibitem

%%% 19
\bibitem[\protect\citeauthoryear{Unger and Farrar}{2017}]{Unger:2017kfh}
\begin{botherref}
\oauthor{\bsnm{Unger}, \binits{M.}},
\oauthor{\bsnm{Farrar}, \binits{G.R.}}:
{Uncertainties in the Magnetic Field of the Milky Way}
(2017)
{\href{https://arxiv.org/abs/1707.02339}{{arXiv:1707.02339}}}
{[astro-ph.GA]}
\end{botherref}
\endbibitem

%%% 20
\bibitem[\protect\citeauthoryear{Schulte et~al.}{2023}]{Schulte:2023iry}
\begin{barticle}
\bauthor{\bsnm{Schulte}, \binits{J.}},
\bauthor{\bsnm{Bister}, \binits{T.}},
\bauthor{\bsnm{Erdmann}, \binits{M.}}:
\batitle{{An all-sky search method for coherent magnetic field deflections of ultra-high-energy cosmic rays based on Deep Learning}}.
\bjtitle{PoS}
\bvolume{ICRC2023},
\bfpage{198}
(\byear{2023})
\doiurl{10.22323/1.444.0198}
\end{barticle}
\endbibitem

%%% 21
\bibitem[\protect\citeauthoryear{Alves~Batista et~al.}{2022}]{AlvesBatista:2022vem}
\begin{barticle}
\bauthor{\bsnm{Alves~Batista}, \binits{R.}}, \betal:
\batitle{{CRPropa 3.2 \textemdash{} an advanced framework for high-energy particle propagation in extragalactic and galactic spaces}}.
\bjtitle{JCAP}
\bvolume{09},
\bfpage{035}
(\byear{2022})
\doiurl{10.1088/1475-7516/2022/09/035}
{\href{https://arxiv.org/abs/2208.00107}{{arXiv:2208.00107}}}
{[astro-ph.HE]}
\end{barticle}
\endbibitem

%%% 22
\bibitem[\protect\citeauthoryear{Halim et~al.}{2024}]{PierreAuger:2023htc}
\begin{barticle}
\bauthor{\bsnm{Halim}, \binits{A.A.}}, \betal:
\batitle{{Constraining models for the origin of ultra-high-energy cosmic rays with a novel combined analysis of arrival directions, spectrum, and composition data measured at the Pierre Auger Observatory}}.
\bjtitle{JCAP}
\bvolume{01},
\bfpage{022}
(\byear{2024})
\doiurl{10.1088/1475-7516/2024/01/022}
{\href{https://arxiv.org/abs/2305.16693}{{arXiv:2305.16693}}}
{[astro-ph.HE]}
\end{barticle}
\endbibitem

%%% 23
\bibitem[\protect\citeauthoryear{Xiong et~al.}{2021}]{xiong2021}
\begin{botherref}
\oauthor{\bsnm{Xiong}, \binits{Y.}},
\oauthor{\bsnm{Zeng}, \binits{Z.}},
\oauthor{\bsnm{Chakraborty}, \binits{R.}},
\oauthor{\bsnm{Tan}, \binits{M.}},
\oauthor{\bsnm{Fung}, \binits{G.}},
\oauthor{\bsnm{Li}, \binits{Y.}},
\oauthor{\bsnm{Singh}, \binits{V.}}:
Nystr\"omformer: A nystr\"om-based algorithm for approximating self-attention
(2021)
{\href{https://arxiv.org/abs/2102.03902}{{arXiv:2102.03902}}}
{[cs.CL]}
\end{botherref}
\endbibitem

%%% 24
\bibitem[\protect\citeauthoryear{G\'orski et~al.}{2005}]{Gorski:2004by}
\begin{barticle}
\bauthor{\bsnm{G\'orski}, \binits{K.M.}},
\bauthor{\bsnm{Hivon}, \binits{E.}},
\bauthor{\bsnm{Banday}, \binits{A.J.}},
\bauthor{\bsnm{Wandelt}, \binits{B.D.}},
\bauthor{\bsnm{Hansen}, \binits{F.K.}},
\bauthor{\bsnm{Reinecke}, \binits{M.}},
\bauthor{\bsnm{Bartelman}, \binits{M.}}:
\batitle{{HEALPix - A Framework for high resolution discretization, and fast analysis of data distributed on the sphere}}.
\bjtitle{Astrophys. J.}
\bvolume{622},
\bfpage{759}--\blpage{771}
(\byear{2005})
\doiurl{10.1086/427976}
{\href{https://arxiv.org/abs/astro-ph/0409513}{{arXiv:astro-ph/0409513}}}
\end{barticle}
\endbibitem

%%% 25
\bibitem[\protect\citeauthoryear{Zonca et~al.}{2019}]{Zonca:2019vzt}
\begin{barticle}
\bauthor{\bsnm{Zonca}, \binits{A.}},
\bauthor{\bsnm{Singer}, \binits{L.}},
\bauthor{\bsnm{Lenz}, \binits{D.}},
\bauthor{\bsnm{Reinecke}, \binits{M.}},
\bauthor{\bsnm{Rosset}, \binits{C.}},
\bauthor{\bsnm{Hivon}, \binits{E.}},
\bauthor{\bsnm{Gorski}, \binits{K.}}:
\batitle{{healpy: equal area pixelization and spherical harmonics transforms for data on the sphere in Python}}.
\bjtitle{Journal of Open Source Software}
\bvolume{4}(\bissue{35}),
\bfpage{1298}
(\byear{2019})
\doiurl{10.21105/joss.01298}
\end{barticle}
\endbibitem

%%% 26
\bibitem[\protect\citeauthoryear{Sundararajan et~al.}{2017}]{sundararajan2017}
\begin{botherref}
\oauthor{\bsnm{Sundararajan}, \binits{M.}},
\oauthor{\bsnm{Taly}, \binits{A.}},
\oauthor{\bsnm{Yan}, \binits{Q.}}:
Axiomatic attribution for deep networks
(2017)
{\href{https://arxiv.org/abs/1703.01365}{{arXiv:1703.01365}}}
{[cs.LG]}
\end{botherref}
\endbibitem

\end{thebibliography}
%%\input attention.bbl


\end{document}